\documentstyle[aps,twocolumn,epsfig]{revtex}
\begin{document}
\title{Spatial mapping of the electron eigenfunctions
in InAs self-assembled quantum dots by magnetotunneling}
\author{E.\,E. Vdovin$^1$, Yu.\,N. Khanin$^1$,  A.\,V.
Veretennikov$^3$, A. Levin$^2$, A. Patane$^2$, Yu.\,V.
Dubrovskii$^1$, L. Eaves$^2$, P.\,C. Main$^2$, M. Henini$^2$, G.
Hill$^4$}
\address{$^1$Institute of Microelectronics
Technology RAS, Chernogolovka, Russia\\ $^2$School of Physics and
Astronomy, University of Nottingham, Nottingham NG7 2RD, UK\\
$^3$Institute of Solid State Physics RAS, Chernogolovka, Russia\\
$^4$Dept. of Electronic and Electrical Engineering, University of
Sheffield, UK}
\date{June 28, 2001}
\maketitle
\begin{abstract}
We use magnetotunnelling spectroscopy
as a non-invasive probe to produce two-dimensional spatial images
of the probability density of an electron confined in a
self-assembled semiconductor quantum dot. The images reveal
clearly the elliptical symmetry of the ground state and the
characteristic lobes of the higher energy states.
\end{abstract}

\section{ Introduction}

Quantum dots (QDs) are characterised by relatively
small number of electrons confined within an island with a
nanometer dimension. They can confine the motion of an electron in
all three spatial dimensions ~\cite{c1}. The strong confinement in
the QD gives rise to a set of discrete and narrow electronic
energy levels similar to those in atomic physics. The epitaxial
growth of lattice mismatched $InAs$ on $GaAs$ or $AlAs$ opens new
possibilities for the simple fabrication of semiconductor
nanostructures. $InAs$ QDs are formed {\it in situ} during growth
due to the relaxation of a strained $InAs$ wetting layer on $GaAs$
or $AlAs$ ~\cite{c2}. The particular interest lies in their
uniformity and small size: a lateral dimension $10-20~nm$  and
height $3-4$ nm. Several theoretical approaches have been used to
calculate the eigenstates of $InAs$ QDs ~\cite{c3}. The results of
the calculation depend strongly on the assumed shape and
composition of the QDs. Experimentally, the quantized energy
levels of a given potential can be probed using various
spectroscopic techniques. The corresponding wave functions are
much more difficult to measure. Information about the extent of
the carrier wavefunction for the ground state of a QD was obtained
from tunneling measurements in a magnetic field ~\cite{c4}. Also,
the anisotropy of electronic wave function in self-aligned $InAs$
QDs was deduced from magnetic-field-dependent photoluminescence
spectroscopy ~\cite{c5}. However, until recently there have been
no reported measurements of the detailed spatial form of the
wavefunctions of the ground and excited states of the QD.
Recently, it has been demonstrated that magnetotunneling
spectroscopy can be employed as nonivasive probe to produce images
of the probability density of the electron confined in a QD
~\cite{c6}. In this work, we use magnetotunneling spectroscopy
(MTS) to investigate in detail of the spatial form of the wave
function of the electron states of a double-barrier resonant
tunnelling diode with $InAs$ QDs embedded in the centre of a
$GaAs$ quantum well. We measure the dependence of the resonant
tunnelling current through the QD states as a function of magnetic
field, B, applied perpendicular to the tunnelling direction. This
allows us to map out the full spatial form of the probability
density of the ground and excited states of the QDs. The electron
wavefunctions have a biaxial symmetry in the growth plane, with
axes corresponding quite closely (within measurement error of
15$\raisebox{1ex}{\scriptsize o}$) to the main crystallographic
directions $X-[01\bar 1]$  and $Y-[\bar 233]$ for (311)B substrate
orientation. For a similar $InAs$ QD structure grown on a
(100)-substrate we also obtained characteristic probability
density maps of ground and exited states.

\section{Experimental details} The $InAs$ QDs are embedded in a
$n-i-n$, resonant tunnelling diode. The samples were grown by
molecular beam epitaxy on a $GaAs$ substrate with the orientations
(100) and (311)B. A layer of $InAs$ QDs, nominally $2.3$ monolayer
thick, was placed in the centre of a $9.6~nm$ wide $GaAs$ quantum
well (QW) with $8.3~nm$ $Al_{0.4}Ga_{0.6}As$ confining barriers,
sandwiched between two nominally-undoped $50~nm$ $GaAs$ spacer
layers. The intrinsic region is surrounded by graded n-type
contact layers, with the doping concentration increasing from $2
\times 10 ^{17} cm^{3}$ to $3 \times 10 ^{18} cm^{-3}$. The layers
were grown at 600$\raisebox{1ex}{\scriptsize o}$ C, and there was
a growth interrupt before the QDs were grown at 480
$\raisebox{1ex}{\scriptsize o}$C.  For comparison, we also studied
two control samples grown with the same sequence of layers, except
one has only a thin $InAs$ two-dimensional wetting layer (i.e.
containing no QDs) and the other has no $InAs$ layer at all. The
samples were processed into circular mesa structures of diameters
between $50~\mu m$ and $200~\mu m$, with ohmic contacts to the
doped regions.

\section{Results and discussion}

\begin{figure}\centering{\epsffile{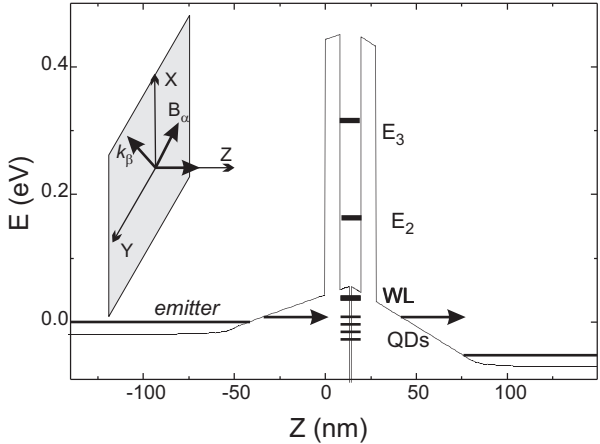}}
\caption{
Schematic conduction band profile under an applied bias
of an n-i-n $GaAs$/($AlGa$)$As$ double barrier resonant tunneling
diode incorporating $InAs$ self-assembled quantum dots (QDs).
Inset: Orientation of the magnetic field, $B$, and current,$I$, in
the magnetotunnelling experiment. X and Y define the two main
crystallographic axis, $[01\bar 1]$  and $[\bar 233]$,
respectively, in the (311)-oriented GaAs substrate. $\alpha $ and
$\beta $ indicate, respectively, the direction of $B$ and of the
momentum acquired by the tunnelling electron due to the action of
the Lorentz force.}
\end{figure}
Figure 1 shows a schematic energy band diagram for our device
under bias voltage. X and Y define the two main crystallographic
axis in the plane perpendicular to growth direction, Z (see
inset). The layer of $InAs$ QDs introduces a set of discrete
electronic states below the $GaAs$ conduction band edge. At zero
bias voltage, equilibrium is established by electrons diffusing
from the doped $GaAs$ layers and filling some dot states. The
resulting negative charge in the QW produces depletion layers in
the regions beyond the (AlGa)As barriers. By applying a bias
voltage to the emitter layer, V, the QD energy level is shifted in
energy with respect to both contacts. When a particular dot state
is resonant with an adjacent filled state in the biased electron
emitter layer, electrons tunnel through the dot into the collector
and a current flows as shown schematically in Fig.1. Therefore, as
we adjust the voltage we can study different energy states of the
QDs. At sufficiently high voltages we are able to observe two
separate resonances in the current related to confined subbands of
the QW states.

\begin{figure}\centering{\epsffile{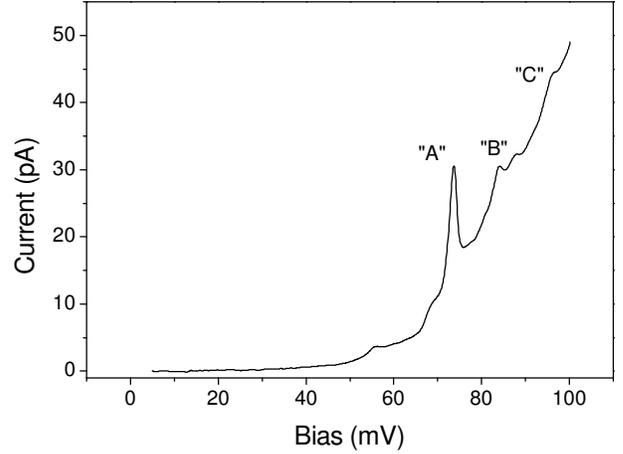}}
\caption{Low-temperature ($T = 1.2~K$) current-voltage
characteristics, $I(V)$. Dots are grown on (311)B substrate
orientation.}
\end{figure}
Figure 2 shows a typical low-temperature ($T=1.2~K$) $I(V)$ curve.
The device contains $InAs$ QDs grown on a (311)B-oriented $GaAs$
substrate. Similar   were obtained for QDs grown on a
(100)-substrate. We observe a series of resonant peaks
corresponding to carrier tunnelling into the dot states.
Pronounced current features appear at biases as low bias as 55 mV.
They are superimposed on a rising background current and cannot be
resolved for $V > 200~mV$. These features are not observed in our
control samples which do not contain QDs and are thereforewe can
ascribe them to the $InAs$ QD layer. Despite the large number of
quantum dots in our sample ($10^6-10^7$ for a $100 \mu m$ diameter
mesa), we observed only a small number of resonant peaks over the
bias range ($\sim 200~mV$) close to the threshold for current
flow. This behaviour has been reported in earlier studies
~\cite{c4,c6,c8,c10,c11,c12,c13}
 and, although not fully understood, is probably related to the limited number of
conducting channels in the emitter that can transmit electrons
from the doping layer to the quantum dots at low bias. There is no
reason to believe that the dots studied are atypical of the
distribution as a whole. On increasing the temperature to $4.2~K$,
the main peaks are still prominent, but much weaker features,
which may be related to density-of-state fluctuation in the
emitter ~\cite{c14}, are strongly suppressed. A key observation is
that many peaks look similar so we cannot tell if the peaks are
due to tunneling through the states of a single dot or several
dots. In the following, we will concentrate on three voltage
regions labelled ("A"), ("B") and ("C"). We will focus on the
magnetic field dependence of the QD resonances and on how this
provides detailed information about the form of the wavefunction
associated with an electron in the QD state.

\begin{figure}\centering{\epsffile{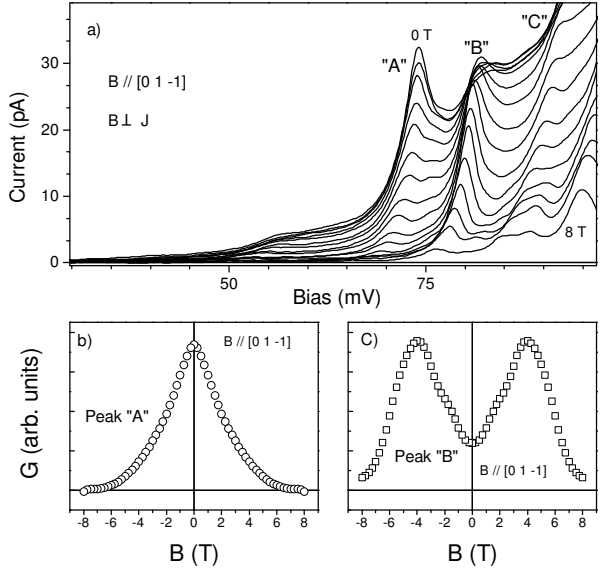}}
\caption{(a) Low-temperature ($T = 4.2~K$) $I(V)$ characteristics
in the presence of a magnetic field, $B$. The direction of $B$ is
perpendicular to the current flow. (b) and (c) Dependence of
conductance, $G$, on magnetic field for $B$ parallel to
$[01\bar1]$ for different QDs states. $B$ was varied from 0 to 8 T
with the step of 0.5 T.}
\end{figure}
Figure 3(a) shows the low-temperature ($T = 4.2~K$) $I(V)$
characteristics in the presence of a magnetic field, $B$. The
direction of $B$ is perpendicular to the current flow and lies in
the (X, Y) plane (see Fig 1). The axis $[01\bar 1]$  and $[\bar
233]$  define the two main crystallographic axis in the plane
perpendicular to growth direction [311]. The amplitude of each
resonance exhibits a strong dependence on the intensity of $B$. In
particular, with increasing $B$, the low-voltage resonances "A"
decrease steadily in amplitude, whereas the others, "B" and "C",
have a non- monotonic magnetic field dependence. The Figure 3(b)
and (c) show clearly two characteristic types of magnetic field
dependence: type "A" shows a maximum on G(B) at B = 0 T followed
by a steady decay to zero at around $8 T$; type "B" shows a broad
maximum at $\sim $ $4.5~T$, followed by a gradual decay to zero.

\begin{figure}\centering{\epsffile{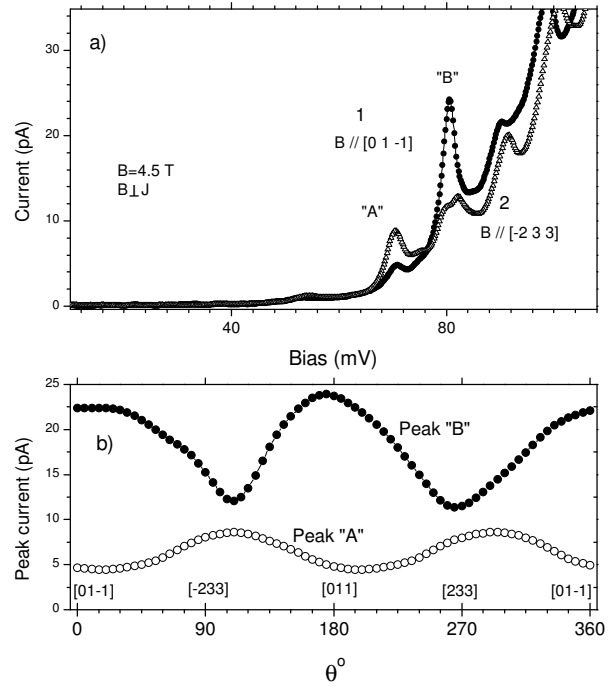}}
\caption{(a) $I(V)$ characteristics in an in-plane magnetic field
of $4.5~T$. The first curve (circles) is for $B\|[01\bar1]$; the
second curve (triangles) is for $B\|[\bar 233]$; (b) angular
dependence of the peaks current.}
\end{figure}
Figure 4(a) shows the $I(V)$ characteristics in an in-plane
magnetic field of $4.5~T$. The first curve (circles) is for
$B\|[01\bar 1]$  ; the second curve (triangles) is for $B\|[\bar
233]$. We observe a clear anisotropy in the dependence of $I(V)$
on $B$ for the two field orientations. As can be clearly seen in
Fig.4(a) peaks "A" and "B" in the $I(V)$ plot reveal a strong
anisotropy of about $\rho \sim$ 0.5. We have also determined
angular dependence of the peaks. The results are plotted in
Fig.4(b) for  peaks "A" and "B". Note that all peaks, observed
over the bias range ($\sim$200 mV) have a maxima  in current
amplitude at orientation of a field $B\|[01\bar 1]$  or $B\|[\bar
233]$. The main effect to be noted from Fig.4 is the dependence of
the current as a function of the in-plane magnetic field
orientation.

We can understand the magnetic field dependence of the features in
terms of the effect of $B$ on a tunnelling electron. Let $\alpha
$, $\beta $, and $Z$ indicate, respectively, the direction of $B$,
the direction normal to $B$ in the growth plane $(X, Y)$, and the
normal to the tunnel barrier, respectively (see Figure 1 (b)).
When an electron tunnels from the emitter into the dot, it
acquires an additional in-plane momentum given by ~\cite{c18}

\begin{equation}
k_{\beta} =\frac{eB \Delta s}{\hbar}
\label{1}
\end{equation}

where $\Delta s$  is the effective distance tunnelled along $Z$.
This effect can be understood semiclassically in terms of the
increased momentum along $\beta$, which is acquired by the
tunnelling electron due to the action of the Lorentz force. In
terms of mapping out the spatial form of an electronic state, we
can envisage the effect of this shift in   as analogous to that of
the displacement, in real space, of the atomic tip in a STM
imaging measurement. The applied voltage allows us to tune
resonantly to the energy of a particular QD state. Then, by
measuring the variation of the tunnel current with $B$, we can
determine the size of the matrix element that governs the quantum
transition of an electron as it tunnels from a state in the
emitter layer into a QD. In our experiment, the tunnelling matrix
element is most conveniently expressed in terms of the Fourier
transforms $\Phi_{i,f} (\underline {k})$ of the conventional real
space wavefunctions ~\cite{c18,c19}. Here the subscripts $i$ and
$f$ indicate the initial (emitter) and final (QD) states of the
tunnel transition. Relative to the strong spatial confinement in
the QD, the initial emitter state has only weak spatial
confinement. Hence, in  \underline {$k$}-space corresponds to a
sharply peaked function with a finite value only close to
\underline {$k$} = 0. Since the tunnel current is given by the
square of the matrix element involving $\Phi_i (\underline {k})$
and $\Phi_{QD} (\underline {k})$, the narrow spread of \underline
{$k$} for $\Phi_i (\underline {k})$ allows us to determine the
form of $\Phi_{QD} (\underline {k})$ by varying $B$ and hence
\underline {$k$} according to (\ref{1}). Thus by plotting $G(B)$
for a {\it particular} direction of $B$ we can measure the
dependence of $|\Phi_{QD} (\underline {k})|^2$ along the
\underline {$k$}-direction perpendicular to $B$. Then, by rotating
$B$ in the plane (X, Y) and making a series of measurements of
$I(B)$ with $B$ set at regular intervals ( $\Delta \theta \sim$
5$\raisebox{1ex}{\scriptsize o}$) of the rotation angle $\theta $,
we obtain a full spatial profile of $|\Phi _{QD} (k_X , k_Y)|^2$.
This represents the projection in \underline {$k$}-space of the
probability density of a given electronic state confined in the
QD.

The model provides a simple explanation of the magnetic field
dependence of the resonant current features "A-C". In particular,
the {\it forbidden} nature of the tunnelling transition associated
with "B" at $B = 0~T$ is due to the odd parity of the final state
wavefunction, which corresponds to the first excited state of a
QD. The applied magnetic field (i.e. the Lorentz force)
effectively breaks the mirror symmetry at $B = 0$ and thus makes
the transition {\it allowed}.

\begin{figure}\centering{\epsffile{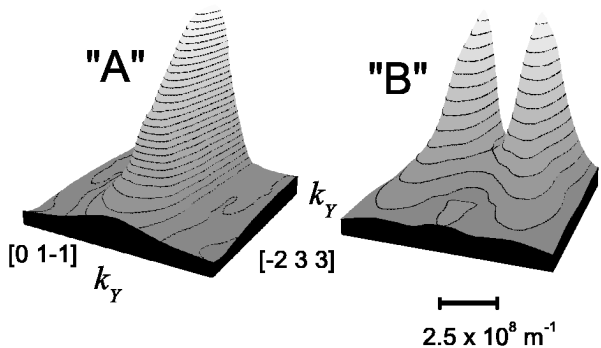}}
\caption{
Distribution in the plane $(k_X , k_Y)$ of the differential
conductance, $G = dI/dV$, for three representative states. This
provides a spatial map of $|\Phi _{QD} (k_X , k_Y)|^2$, the square
of the Fourier transform, $\Phi _{QD} (k_X , k_Y)$, of the
probability density of the electron confined in the dot. $X$ and
$Y$ define the two main crystallographic axis, $[01\bar1]$ and
$[\bar233]$, respectively, in the (311)-oriented $GaAs$ plane.}
\end{figure}
Figure 5 shows the spatial form of $G(B) \sim |\Phi _{QD} (k_X ,
k_Y)|^2$ , in the plane $(k_X , k_Y)$ for the two representative
QD states corresponding to the labelled features in Fig.3(b) and
(c). The  \underline {$k$}-values are estimated from relation (1),
assuming $\Delta s$  has nominal value of $30$~nm which we
estimate from capacitance measurements and from the doping profile
and composition of the device. The contour plots reveal clearly
the characteristic form of the probability density distribution of
a ground state orbital and the characteristic lobes of the higher
energy states of the QD. The electron wavefunctions have a biaxial
symmetry in the growth plane, with axes corresponding quite
closely (within measurement error of 15$\raisebox{1ex}{\scriptsize
o}$) to the main crystallographic directions $X-[01\bar 1]$  and
$Y-[\bar 233]$  for (311) B-substrate orientation. For a similar
$InAs$ QD structure grown on a (100)-substrate we also obtained
characteristic probability density maps of ground and exited
states.

To summarise, we have observed features in $I(V)$ corresponding to
resonant tunnelling through a limited number of discrete states
whose wavefunctions display the symmetry of the ground state and
excited states of quantum dots. With the simple device
configuration we have used, it is not possible to tell whether an
excited state feature and a ground state feature correspond to the
same quantum dot. This question could be resolved by new
experiments on structures with electrostatic gates.

\section{Conclusion} In conclusion, we have shown how the
magnetotunnelling spectroscopy provides a new means of probing the
spatial form of the wavefunctions of electrons confined in quantum
dots. The study revealed a biaxial symmetry of QD states in the
growth plane. We observed the elliptical shape of the ground state
and the characteristic lobes of the higher energy states.

\section*{Acknowledgements} The work is partly supported by RFBR
(00-02-17903), INTAS-RFBR(95-849) and EPSRC (UK). AL gratefully
acknowledges the support of the FAPESP Foundation (Brazil). EEV
gratefully acknowledge supports from the Royal Society. We are
gratefully to V.A.Tulin and V.G. Lysenko for helpful discussions
and V.V. Belov, A. Orlov and D.Yu. Ivanov for technical
assistance.

\end{document}